\documentclass[11pt,dvips]{article}
\usepackage{eurosym}
\usepackage{graphicx}
\usepackage{amsmath}
\usepackage{amssymb}
\usepackage{latexsym}
\usepackage{color}
\usepackage{epstopdf}

\input{tcilatex}
\begin{document}

\thispagestyle{empty}

\begin{center}
\vspace{0.7cm}

%%%%%%%%%%%%%%%%%%%%%%%%%%%%%%%%%%%%%%%%%%%%%%%%%%%%%%%%%%%%%%%%%%%%%%%%%%%%%%%%%%%%%%%%%%%%%%%%%%%%%%%%%%%%%%%%%%%%%%%
{\Large \textbf{Gaussian R\'{e}nyi-2 correlations in a nondegenerate
three-level laser}}%
%%%%%%%%%%%%%%%%%%%%%%%%%%%%%%%%%%%%%%%%%%%%%%%%%%%%%%%%%%%%%%%%%%%%%%%%%%%%%%%%%%%%%%%%%%%%%%%%%%%%%%%%%%%%%%%%%%%%%

\bigskip \textbf{Jamal El Qars}

\textit{Laboratory of Materials, Electrical Systems, Energy and Environment
(LMS3E)}

\textit{Faculty of Applied Sciences, Ait-Melloul, Ibn Zohr University,
Agadir, Morocco}\vspace{0.07cm}

\vspace{0.9cm}

\vspace{0.9cm}\textbf{Abstract}
\end{center}

Quantum correlation is a key component in various quantum information
processing tasks. Decoherence process imposes limitations on achieving these
quantum tasks. Therefore, understanding the behavior of quantum correlations
in dissipative-noisy systems is of paramount importance. Here, on the basis
of the Gaussian R\'{e}nyi-2 entropy, we analyze entanglement and quantum
discord in a two-mode Gaussian state $\rho_{AB}$. The mode $A$($B$) is
generated within the first(second) transition of a nondegenerate three-level
cascade laser. Using realistic experimental parameters, we show that both
entanglement and discord could be generated and enhanced by inducing more
quantum coherence. Under thermal noise, entanglement is found more fragile
having a tendency to disappear rapidly. While, quantum discord exhibits a
freezing behavior, where it can be captured within a wide range of
temperature. Surprisingly, we find that entanglement can exceed quantum
discord in contrary to the expectation based on the assumption that the
former is only a part of the later. Finally, we show numerically as well as
analytically that optimal quantum discord can be captured by performing
Gaussian measurements on mode $B$. The obtained results suggest that
nondegenerate three-level lasers may be a valuable resource for some quantum
information tasks, especially, for those do not require entanglement.

\section{Introduction}

Quantum systems are correlated in manners inaccessible to classical ones
\cite{Nielson}. A peculiar quantum characteristic of correlations is quantum
entanglement \cite{Werner}, which was defined as a nonclassical physical
property that cannot be prepared by means of local operations and classical
communication \cite{Werner}. Entanglement is undoubtedly the key ingredient in most applications of
quantum science \cite{Horodecki}, where it has been recognized as the
fundamental resource for, e.g., quantum teleportation \cite{Luo}, dense
coding \cite{Chen}, quantum computation \cite{div} and quantum algorithm
\cite{Harow}. However, it is shown that some quantum tasks, e.g., quantum
key distribution \cite{Grosshans} can be carried out by unentangled states
that nevertheless possess quantum correlations \cite{laflamme}. In fact, it has been proven theoretically \cite{laflamme} as well as
experimentally \cite{ex1} that some quantum tasks may be speed-up over their
classical counterparts exploiting unentangled states with residual
correlations that cannot be described by any classical probability
distribution. Such residual correlations are quantified by the so-called
quantum discord \cite{Modi}.

For bipartite quantum states endowed with finite-dimensional Hilbert spaces,
the concept of discord was first introduced and defined as the mismatch
between total and classical correlations \cite{Zurek,Anderson}. The
definition of quantum discord involves a nontrivial optimization task which
can be accomplished solely for very simple states, including X-states \cite%
{mazhar} and two-mode Gaussian states \cite{GS}. Restricting the
minimization---implicated in the definition of quantum discord \cite%
{Zurek,Anderson}---to the set of Gaussian positive operator-valued measures
\cite{Cirac}, the optimization problem has been fully solved in \cite%
{Adesso,ASG}. Quantum discord was predicted to play the main role in miscellaneous
protocols, e.g., quantum state merging \cite{merging}, remote state
preparation \cite{rsp}, security in quantum key distribution \cite{scar} and
quantum channel discrimination \cite{qsd}. In this regard, quantum discord
has been investigated in different systems including two-qubit states \cite%
{mazhar,Luo2}, two resonant harmonic oscillators \cite{Paz}, photonic
crystal cavity array \cite{oh}, optomechanical Fabry-P\'{e}rot cavities \cite%
{Joptic}. The experimental exploration of quantum discord is accomplished in
\cite{ex1}.

Notice that steering \cite{UolaGuhne} and Bell nonlocality \cite{Bell} are
also two other incarnations of quantum correlations that can be, especially,
used to implement secure quantum information processing tasks, e.g, quantum
secret sharing protocol \cite{QSS} and unconditionally secure quantum key
distribution \cite{Gisin}. Another fundamental aspect that marks the
departure of the quantum science from the classical one is quantum coherence
\cite{coherence}. Such purely quantum property constitutes a powerful
resource for quantum information processing \cite{coherence,Bromly}, and
plays a fundamental role in emergent fields like quantum biology \cite{SSP}
and thermodynamics \cite{Thermo}. In recent years, the topic of Gaussian quantum correlations has received a
significant amount of attention as it plays a crucial role in quantum
computation and communication protocols \cite{GS}. While the efficiency of
quantum information schemes is strongly depends on the degree of quantum
correlations, three-level lasers have been theoretically predicted to be a
good candidate as a source of light in a highly entangled state \cite{SZ}. Scully and Zubairy \cite{SZ} have established the basic tools of a really complete theory of two-photon laser emitted by three-level atoms in a
cascade configuration. In such lasers, the major role is played by the
atomic coherence \cite{SZ}, which can be induced either by the injected
coherence process \cite{Wodkiewicz} or by the driven coherence process \cite%
{ansari1}. Importantly, two-mode light generated within the cascade
transition of a three-level laser is proven to evolve in a two-mode Gaussian
state \cite{tangtan}. A two-photon laser, with a gain media constituted by a
set of three-level atoms in a cascade configuration, is shown to display
several quantum effects such as quenching of quantum fluctuations \cite%
{Kapale}, quantum squeezing effect \cite{tesf74}, as well as anomalous
optical bistability \cite{josa15}.

Over the past 2 decades, miscellaneous works focused only on inseparable
quantum correlations (steering, entanglement, and Bell nonlocality) in
three-level lasers coupled to either vacuum or squeezed reservoirs. For
instance, Ping \textit{et al}. \cite{ping} and Alebachew \cite{Alebachew}
have studied Bell nonlocality, where the atomic coherence is induced via the
driven coherence process and the injected coherence process, respectively.
They found that violation of Bell's inequality of the entangled states is
possible even in the presence of cavity losses. Recently, El Qars \cite{JDER}
has investigated Gaussian quantum steering, where the atomic coherence is
induced by initially preparing the three-level atoms in a coherent
superposition of the upper and lower levels. It is found that due the
positivity of the intensity difference of the two emitted laser modes,
one-way steering behavior can be detected only in one direction. Opposed to
these works, the entanglement properties in three-level lasers have been
extensively examined. Proposals include, but not limited to, \cite%
{tangtan,tesf74,xiong,Bacha,tesfa23}, where the sufficient and necessary
inseparability criterion proposed in \cite{Simon,Duan} is employed as an
entanglement witness.

In realistic quantum systems, quantum correlations are inevitably effected
by the surrounding environment which leads generally to their degradation
\cite{Paz}. This is a challenging issue for generating and preserving
quantum correlations in dissipative-noisy quantum optical systems, which are
of great importance for quantum information processing \cite{Nielson}. Due to the increasing interest in the quantum properties of correlated
emission lasers, we propose here to investigate, against decoherence effect,
both entanglement and discord in a nondegenerate three-level laser where the
atomic coherence is initially induced by the injected coherence process. To
this aim, we consider a two-mode Gaussian state $\rho _{AB}$ coupled to a
common two-mode thermal reservoir. The modes $A$ and $B$ are generated,
respectively, during the first and second transitions of a single
three-level atom. We use the Gaussian R\'{e}nyi-2 entanglement and the
Gaussian R\'{e}nyi-2 discord to quantify, respectively, entanglement and
quantum discord between the two laser modes $A$ and $B$. Finally, we
emphasize that entanglement and Gaussian quantum discord defined via the R%
\'{e}nyi-2 entropy have been studied in an optomechanical system subjects to
dissipation and thermal noise by El Qars \textit{et al}. \cite{Joptic}.

The remainder of this paper is organized as follows. In Sect. \ref{s2}, we
introduce the system under consideration. Next, by applying the master
equation governing the dynamics of the state $\rho _{AB}$, we derive the
explicit expression of the stationary covariance matrix fully describing the
two-mode Gaussian state $\rho _{AB}$. In Sect. \ref{s3}, using realistic
experimental parameters, we quantify and study the Gaussian R\'{e}nyi-2
entanglement and the Gaussian R\'{e}nyi-2 discord in the state $\rho _{AB}$.
Finally, in Sect. \ref{s4}, we draw our conclusions.

\section{A nondegenerate three-level cascade laser}

\label{s2}

\begin{figure}[tbh]
\centerline{\includegraphics[width=10.5cm]{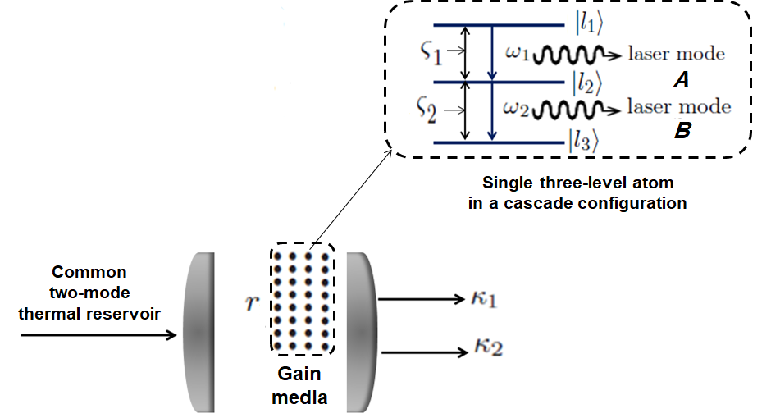}}
\caption{A three-level laser coupled to a common two-mode thermal reservoir.
We adopt the notations $|l_{1}\rangle $, $|l_{2}\rangle $ and $|l_{3}\rangle
$ for representing, respectively, the upper, intermediate, and ground levels
for a single three-level atom. The first(second) transition $|l_{1}\rangle
\rightarrow |l_{2}\rangle $($|l_{2}\rangle \rightarrow |l_{3}\rangle $) at
the optical frequency $\protect\omega _{1}$($\protect\omega _{2}$) and
spontaneous emission decay rate $\protect\gamma _{12}$($\protect\gamma _{23}$%
), is assumed to be in resonance with the quantized cavity mode $\protect%
\varsigma _{1}$($\protect\varsigma _{2}$). While, the transition $%
|l_{1}\rangle \rightarrow |l_{3}\rangle $ is dipole forbidden \protect\cite%
{Fesseha}. The gain media of the laser is constituted by an ensemble of
three-level atoms in a cascade configuration. We focus on the situation in
which the atoms are initially prepared in a coherent superposition of the
upper excited level $|l_{1}\rangle $ and the ground level $|l_{3}\rangle $. $%
r$ denotes the rate at which the atoms are placed in the cavity. When a
single atom makes a transition from the top level $|l_{1}\rangle $ to the
bottom level $|l_{3}\rangle $ via the intermediate level $|l_{2}\rangle $,
two strongly correlated photons are emitted with the frequencies $\protect%
\omega _{1}$ and $\protect\omega _{2}$. If $\protect\omega _{1}\neq\protect%
\omega _{2}$, which we consider here, the laser is a nondegenerate
three-level laser, and a degenerate three-level laser if $\protect\omega %
_{1}=\protect\omega _{2}$.}
\label{Fig1}
\end{figure}

Inside a resonant cavity, we consider a set of three-level atoms in
interaction with two bosonic modes of the quantized cavity radiation \cite%
{SZ}. The $j$\textrm{th} bosonic mode can be characterized by its
annihilation operator $\varsigma _{j}$, decay rate $\kappa _{j}$ and
frequency $\omega _{j}$. We suppose that the atomic system is injected in
the cavity with a rate $r$ \cite{Fesseha}. As illustrated in Fig. \ref{Fig1}%
, the notations $|l_{1}\rangle ,$ $|l_{2}\rangle $ and $|l_{3}\rangle $ are
used to indicate, respectively, the upper excited, intermediate and ground
levels of a single three-level atom.

The interaction between the two cavity modes $\varsigma _{1}$ and $\varsigma
_{2}$ and a single atom can be described by the Hamiltonian \cite{SZ}
\begin{equation}
\mathcal{H}_{\mathrm{int}}=\mathrm{i}\hbar \left[ \upsilon _{12}\varsigma
_{1}|l_{1}\rangle \langle l_{2}|+\upsilon _{23}\varsigma _{2}|l_{2}\rangle
\langle l_{3}|-\upsilon _{12}|l_{2}\rangle \langle l_{1}|\varsigma
_{1}^{\dag }-\upsilon _{23}|l_{3}\rangle \langle l_{2}|\varsigma _{2}^{\dag }%
\right] ,  \label{E1}
\end{equation}%
where $\upsilon _{12}$ are $\upsilon _{23}$ being the coupling constants
corresponding to the first $|l_{1}\rangle \rightarrow |l_{2}\rangle $ and
second$\ |l_{2}\rangle \rightarrow |l_{3}\rangle $ transitions, respectively
\cite{Fesseha}. In addition, we suppose that the three-level atoms are
initially prepared in an arbitrary quantum coherent superposition of the
excited upper level $|l_{1}\rangle $ and the ground level $|l_{3}\rangle $.
Therefore, the initial state $|\Psi _{\mathrm{sa}}\rangle $ as well as the
density operator $\rho _{\mathrm{sa}}$ for a single atom read \cite{CS}
\begin{eqnarray}
|\Psi _{\mathrm{sa}}\rangle &=&p_{1}|l_{1}\rangle +p_{3}|l_{3}\rangle ,
\label{E2} \\
\rho _{\mathrm{sa}} &=&\rho _{11}|l_{1}\rangle \langle l_{1}|+\rho
_{13}|l_{1}\rangle \langle l_{3}|+\rho _{31}|l_{3}\rangle \langle
l_{1}|+\rho _{33}|l_{3}\rangle \langle l_{3}|,  \label{E3}
\end{eqnarray}%
where $\rho _{11}=|p_{1}|^{2}$ and $\rho _{33}=|p_{3}|^{2}$ represent,
respectively, the probabilities for a single three-level atom to be
initially in the excited upper level and the ground level. While $\rho
_{13}=\rho _{31}^{\ast }=p_{1}p_{3}^{\ast }$ is the initial coherence of a
single three-level atom \cite{Fesseha}. For simplicity, we take the same
spontaneous decay rates for the transitions $|l_{1}\rangle \rightarrow
|l_{2}\rangle $ and $|l_{2}\rangle \rightarrow |l_{3}\rangle $, i.e., $%
\gamma _{12,23}=\gamma $, the same coupling transitions, i.e., $\upsilon
_{12,23}=\upsilon $, and the same damping rates, i.e., $\kappa _{1,2}=\kappa
.$

The dynamics of the reduced density operator $\rho _{AB}$ for the two laser
modes $A$ and $B$---emitted during the first and second transitions,
respectively---is described by the master equation \cite{Louissel,Akramine}
\begin{equation}
\frac{d\rho _{AB}}{dt}=\frac{-\text{i}}{\hbar }\mathrm{Tr}_{\mathrm{sa}}[%
\mathcal{H}_{\mathrm{int}},\rho _{\{\mathrm{sa+}AB\}}]+\sum\limits_{j=1,2}%
\left( \frac{\kappa _{j}(n_{\mathrm{th},j}+1)}{2}\mathcal{L}\left[ \varsigma
_{j}\right] \rho _{AB}+\frac{\kappa _{j}n_{\mathrm{th},j}}{2}\mathcal{L}%
[\varsigma _{j}^{\dag }]\rho _{AB}\right) ,  \label{E4}
\end{equation}%
with $\rho _{\{\mathrm{sa+}AB\}}$ being the density operator describing the
two laser modes $A$ and $B$ together with a single three-level atom, and $%
\text{Tr}_{\text{sa}}$ denotes the partial trace over the subsystem
constituted by a single atom. In Eq. (\ref{E4}), the Lindblad operator $%
\mathcal{L}\left[ \varsigma _{j}\right] \rho_{AB}=2\varsigma _{j}\rho
_{AB}\varsigma _{j}^{\dag }-\left[ \varsigma _{j}^{\dag }\varsigma _{j},\rho
_{AB}\right] _{+}$ is added to take into account the coupling between the $j%
\text{th}$ cavity mode and the $j\text{th} $ thermal bath having mean
thermal phonon number $n_{\text{th},j}$ \cite{SZ}.

Applying the linear-adiabatic approximation \cite{Sargent} in the good
cavity limit, i.e., $\kappa \ll \gamma $ \cite{SZ} with common thermal bath,
Eq. (\ref{E4}) would be \cite{JDER,JCTP}
\begin{eqnarray}
\frac{d\rho _{AB}}{dt} &=&\frac{\kappa (n_{\mathrm{th}}+1)}{2}[2\varsigma
_{1}\rho _{AB}\varsigma _{1}^{\dag }-\varsigma _{1}^{\dag }\varsigma
_{1}\rho _{AB}-\rho _{AB}\varsigma _{1}^{\dag }\varsigma _{1}]+\frac{\kappa
n_{\mathrm{th}}}{2}[2\varsigma _{2}^{\dag }\rho _{AB}\varsigma
_{2}-\varsigma _{2}\varsigma _{2}^{\dag }\rho _{AB}-\rho _{AB}\varsigma
_{2}\varsigma _{2}^{\dag }]+  \notag \\
&&\frac{1}{2}\left( \mathcal{A}\rho _{11}+\kappa n_{\mathrm{th}}\right)
[2\varsigma _{1}^{\dag }\rho _{AB}\varsigma _{1}-\varsigma _{1}\varsigma
_{1}^{\dag }\rho _{AB}-\rho _{AB}\varsigma _{1}\varsigma _{1}^{\dag }]+
\notag \\
&&\frac{1}{2}\left( \mathcal{A}\rho _{33}+\kappa (n_{\mathrm{th}}+1)\right)
[2\varsigma _{2}\rho _{AB}\varsigma _{2}^{\dag }-\varsigma _{2}^{\dag
}\varsigma _{2}\rho _{AB}-\rho _{AB}\varsigma _{2}^{\dag }\varsigma _{2}]+
\notag \\
&&\frac{\mathcal{A}\rho _{13}}{2}[\rho _{AB}\varsigma _{1}^{\dag }\varsigma
_{2}^{\dag }-2\varsigma _{1}^{\dag }\rho _{AB}\varsigma _{2}^{\dag
}+\varsigma _{1}^{\dag }\varsigma _{2}^{\dag }\rho _{AB}-2\varsigma _{2}\rho
_{AB}\varsigma _{1}+\varsigma _{1}\varsigma _{2}\rho _{AB}+\rho
_{AB}\varsigma _{1}\varsigma _{2}],  \label{E5}
\end{eqnarray}%
with $\mathcal{A}=2r\upsilon ^{2}/\gamma ^{2}$ being the linear gain
coefficient that quantifies the rate at which the atoms are injected into
the cavity \cite{SZ}. In Eq. (\ref{E5}), the term proportional to $\rho
_{33} $($\rho _{11}$) represents the losses(gain) of the mode $B$($A$),
while that proportional to $\rho _{13}$ represents the coupling between the
two modes $A $ and $B$ \cite{SZ}.

Now, utilizing Eq. (\ref{E5}) and the formula $\langle \frac{d\mathcal{O}}{dt%
}\rangle =\mathrm{Tr}\left[ \left( \frac{d\rho _{AB}}{dt}\right) \mathcal{O}%
\right] $, we get the dynamics of the first and second moments of the
variables associated with the laser modes $A$ and $B$, i.e.,
\begin{eqnarray}
\frac{d}{dt}\langle \varsigma _{j}\rangle &=&\frac{-\emptyset _{j}}{2}%
\langle \varsigma _{j}\rangle +\frac{(-1)^{j}\mathcal{A}\rho _{13}}{2}%
\langle \varsigma _{3-j}^{\dag }\rangle \text{ for }j=1,2,  \label{E6} \\
\frac{d}{dt}\langle \varsigma _{j}^{2}\rangle &=&-\emptyset _{j}\langle
\varsigma _{j}^{2}\rangle +(-1)^{j}\mathcal{A}\rho _{13}\langle \varsigma
_{1}^{\dag }\varsigma _{2}\rangle ,  \label{E7} \\
\frac{d}{dt}\langle \varsigma _{j}^{\dag }\varsigma _{j}\rangle
&=&-\emptyset _{j}\langle \varsigma _{j}^{\dag }\varsigma _{j}\rangle +\frac{%
(-1)^{j}\mathcal{A}\rho _{13}}{2}\left[ \langle \varsigma _{1}^{\dag
}\varsigma _{2}^{\dag }\rangle +\langle \varsigma _{1}\varsigma _{2}\rangle %
\right] +(2-j)\mathcal{A}\rho _{11}+\kappa n_{\mathrm{th}},  \label{E8} \\
\frac{d}{dt}\langle \varsigma _{1}\varsigma _{2}\rangle &=&-\frac{\emptyset
_{1}+\emptyset _{2}}{2}\langle \varsigma _{1}\varsigma _{2}\rangle +\frac{%
\mathcal{A}\rho _{13}}{2}\left[ \langle \varsigma _{1}^{\dag }\varsigma
_{1}\rangle -\langle \varsigma _{2}^{\dag }\varsigma _{2}\rangle +1\right] ,
\label{E9} \\
\frac{d}{dt}\langle \varsigma _{1}\varsigma _{2}^{\dag }\rangle &=&-\frac{%
\emptyset _{1}+\emptyset _{2}}{2}\langle \varsigma _{1}\varsigma _{2}^{\dag
}\rangle +\frac{\mathcal{A}\rho _{13}}{2}\left[ \langle \varsigma
_{1}^{2}\rangle -\langle \varsigma _{2}^{\dag 2}\rangle \right] ,
\label{E10}
\end{eqnarray}%
where $\emptyset _{1}=\kappa -\mathcal{A}\rho _{11}$ and $\emptyset
_{2}=\kappa +\mathcal{A}\rho _{33}.$

By introducing the population inversion $\eta $ defined by $\rho
_{11}=(1-\eta )/2$ with $-1\leqslant \eta \leqslant 1$ \cite{SZ}, and using
both $\rho _{11}+\rho _{33}=1$ and $|\rho _{13}|=\sqrt{\rho _{11}\rho _{33}}$
we get $\rho _{33}=$ $(1+\eta )/2$ and $\rho _{13}=\sqrt{1-\eta ^{2}}/2$.

Finally, by using the steady-state condition, i.e., $\frac{d\langle .\rangle
}{dt}=0$ in Eqs. (\ref{E6})-(\ref{E10}), we get the non-zero correlations%
\begin{eqnarray}
\langle \varsigma _{1}^{\dag }\varsigma _{1}\rangle &=&\frac{n_{\mathrm{th}%
}\left( \eta +1\right) }{2\eta ^{2}}+\frac{\left[ \mathcal{A}\eta \left(
\eta -1\right) +2\kappa n_{\mathrm{th}}\right] \left( 1-\eta \right) }{%
4\left( \kappa +\mathcal{A}\eta \right) \eta ^{2}}+\frac{\left( \eta
^{2}-1\right) \left( 4\kappa n_{\mathrm{th}}-\mathcal{A}\eta \right) }{%
2\left( 2\kappa +\mathcal{A}\eta \right) \eta ^{2}},  \label{E11} \\
\langle \varsigma _{2}^{\dag }\varsigma _{2}\rangle &=&-\frac{n_{\mathrm{th}%
}\left( \eta -1\right) }{2\eta ^{2}}+\frac{\left[ \mathcal{A}\eta \left(
\eta -1\right) +2\kappa n_{\mathrm{th}}\right] \left( 1+\eta \right) }{%
4\left( \kappa +\mathcal{A}\eta \right) \eta ^{2}}+\frac{\left( \eta
^{2}-1\right) \left( 4\kappa n_{\mathrm{th}}-\mathcal{A}\eta \right) }{%
2\left( 2\kappa +\mathcal{A}\eta \right) \eta ^{2}},  \label{E12} \\
\langle \varsigma _{1}\varsigma _{2}\rangle &=&\frac{n_{\mathrm{th}}\sqrt{%
1-\eta ^{2}}}{2\eta ^{2}}+\frac{\left[ \mathcal{A}\eta \left( \eta -1\right)
+2\kappa n_{\mathrm{th}}\right] \sqrt{1-\eta ^{2}}}{4\left( \kappa +\mathcal{%
A}\eta \right) \eta ^{2}}-\frac{\sqrt{1-\eta ^{2}}\left( 4\kappa n_{\mathrm{%
th}}-\mathcal{A}\eta \right) }{2\left( 2\kappa +\mathcal{A}\eta \right) \eta
^{2}},  \label{E13}
\end{eqnarray}%
which are physically meaningful only if $\eta \geqslant 0$, then $0\leqslant
\eta \leqslant 1$. The case $\eta =0$ or equivalently $\rho _{11}=\rho
_{33}= $ $\rho _{13}=1/2$ corresponds to maximum injected initial atomic
coherence in the cavity. While, for $\eta =1$, we have $\rho _{11}=\rho
_{13}=0$ and $\rho _{33}=1$, thus no initially atomic coherence are injected.

As mentioned above, it has been demonstrated in \cite{tangtan} that the
two-photon light generated by a nondegenerate three-level laser evolves in a
two-mode Gaussian state, so, the two-mode Gaussian state $\rho _{AB}$ can be
described by means of its covariance matrix defined as $\left[ \mathcal{V}%
_{AB}\right] _{jj^{\prime }}$ $=\langle \mathcal{U}_{j}\mathcal{U}%
_{j^{\prime }}+\mathcal{U}_{j^{\prime }}\mathcal{U}_{j}\rangle /2$, where $%
\mathcal{U}^{\text{\textrm{T}}}=(\mathcal{X}_{1}\mathcal{,Y}_{1}\mathcal{,X}%
_{2}\mathcal{,Y}_{2})$, with $\mathcal{X}_{j}=(\varsigma _{j}^{\dag
}+\varsigma _{j})/\sqrt{2}$ and $\mathcal{Y}_{j}=\mathrm{i}(\varsigma
_{j}^{\dag }-\varsigma _{j})/\sqrt{2}$. Based on the results given by Eqs. [(%
\ref{E11})-(\ref{E13})], the covariance matrix $\mathcal{V}$ can be
rewritten as
\begin{equation}
\mathcal{V}_{AB}=\left(
\begin{array}{cc}
\mathcal{V}{_{A}} & \mathcal{V}{_{A/B}} \\
\mathcal{V}{_{A/B}^{\mathrm{T}}} & \mathcal{V}{_{B}}%
\end{array}%
\right) ,  \label{E14}
\end{equation}%
where $\mathcal{V}_{A}=a$\mbox{$1 \hspace{-1.0mm}
{\bf l}$}$_{2}$, $\mathcal{V}_{B}=b$\mbox{$1 \hspace{-1.0mm}
{\bf l}$}$_{2}$, and $\mathcal{V}_{A/B}=\mathrm{diag}(c,c^{\prime })$, which
corresponds to an asymmetric two-mode squeezed thermal state \cite{Adesso},
with $a=\langle \varsigma _{1}^{\dag }\varsigma _{1}\rangle +1/2$, $%
b=\langle \varsigma _{2}^{\dag }\varsigma _{2}\rangle +1/2$ and $%
c=-c^{\prime }=\langle \varsigma _{1}\varsigma _{2}\rangle $. The
submatrices $\mathcal{V}_{A}$ and $\mathcal{V}_{B}$ describe the two laser
modes $A$ and $B$, respectively, while $\mathcal{V}_{A/B}$ describes the
correlations between them.

\begin{figure}[t]
\centerline{\includegraphics[width=0.45%
\columnwidth,height=5.5cm]{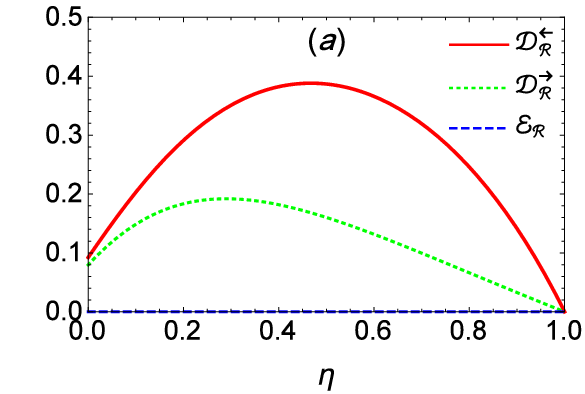}\includegraphics[width=0.45%
\columnwidth,height=5.5cm]{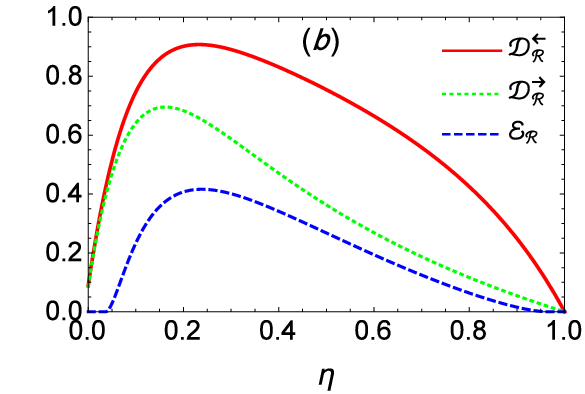}} \centerline{%
\includegraphics[width=0.45\columnwidth,height=5.5cm]{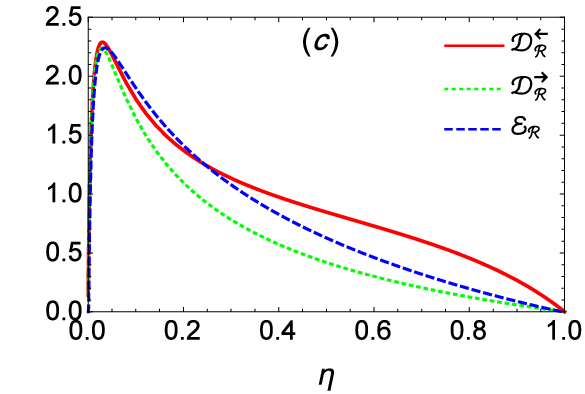}}
\caption{Plot of the Gaussian R\'{e}nyi-2 entanglement $\mathcal{E}_{%
\mathcal{R}}$ and the Gaussian R\'{e}nyi-2 discords $\mathcal{D}_{\mathcal{R}%
}^{\leftarrow}$ and $\mathcal{D}_{\mathcal{R}}^{\rightarrow}$ against the
population inversion $\protect\eta $ using $n_{\mathrm{th}}=5$ as value of
the mean thermal phonon number and $\protect\kappa =3.85~\text{kHz}$ as
value of the cavity decay rate. For the linear gain coefficient, we used $%
\mathcal{A}=100~\text{kHz}$ in panel (a), $\mathcal{A}=1~\text{MHz}$ in
panel (b), and $\mathcal{A}=50~\text{MHz}$ in panel (c). Remarkably, $%
\mathcal{E}_{\mathcal{R}}=\mathcal{D}_{\mathcal{R}}^{\leftarrow}=\mathcal{D}%
_{\mathcal{R}}^{\rightarrow}=0$ for $\protect\eta =1$ regardless of the
values of $\mathcal{A}$. However, optimal entanglement and quantum discord
could be achieved by preparing the three-level atoms in a relatively
stronger coherent superposition ($\mathcal{A}\gg1$ with $\protect\eta %
\rightarrow 0$). Panel (a) shows that quantum discord could be manifested
without entanglement, while panel (c) shows that entanglement may exceed
discord, which clearly indicates that quantum discord can not be viewed as a
sum of entanglement and some other nonclassical correlations.}
\label{Fig2}
\end{figure}

Equations (\ref{E13}) and (\ref{E14}) show that if $\mathcal{A}=0$ or $\eta
=1$, $\det \mathcal{V}_{A/B}=-c^{2}=0$, which turns $\rho _{AB}$ into a
Gaussian product state, i.e., $\rho _{AB}=\rho _{A}\otimes \rho _{B}$
without any correlations (quantum and classical) \cite{Adesso}.
Consequently, neither entanglement nor discord could be created between the
modes $A$ and $B$. This because $\det \mathcal{V}_{A/B}<0$ is a necessary
condition required for the inseparability of the bi-mode Gaussian state $%
\rho _{AB}$ \cite{Simon}. Next, using the Gaussian R\'{e}nyi-2 entropy, we study two different kinds
of quantum correlations, i.e., entanglement and quantum discord in the
cavity radiation of the studied nondegenerate three-level laser.

\section{Gaussian R\'{e}nyi-2 quantum correlations\label{s3}}

\subsection{Gaussian R\'{e}nyi-2 entropy}

The set of R\'{e}nyi-$\alpha $ entropies have been first introduced by
Alfred R\'{e}nyi to generalize the concept of entropy \cite{Renyi}. These
entropies encompass the usual entropic measures, i.e., the Shannon entropy
and the von Neumann entropy \cite{VIDRAL}. Mathematically, the R\'{e}nyi-$%
\alpha $ entropies of a quantum state $\rho $ read as \cite{ASG}
\begin{equation}
\mathcal{S}_{\alpha }\mathcal{(\rho )=}\frac{1}{1-\alpha }\ln \left[ \mathrm{%
Tr}\left( \rho ^{\alpha }\right) \right] \text{ with }\alpha \in (0,1)\cup
(1,+\infty ).  \label{E15}
\end{equation}

The $\mathcal{S}_{\alpha }\mathcal{(\rho )}$-entropies satisfy the following
series of important mathematical properties: they are continuous, invariant
under the unitary operations, and additive on tensor-product states \cite%
{ASG}. Notice that the R\'{e}nyi $\alpha$-entropies $\mathcal{S}_{\alpha }%
\mathcal{(\rho )}$ of discrete probability distributions are always
positive. However, they can be negative in continuous case \cite{Alexey}. In the limit $\alpha \rightarrow 1,$ the entropies $\mathcal{S}_{\alpha }%
\mathcal{(\rho )}$ reduce to the conventional von Neumann entropy $\mathcal{%
S(\rho )=}-\mathrm{Tr}\left( \rho \ln \rho \right) $ \cite{VIDRAL}. We
notice that such entropy is intensively used in various fields of sciences
as well as in quantum information theory (see \cite{Joptic} for more
details).

In pure quantum bipartite states, the von Neumann entropy---commonly known
as the entropy of entanglement---is the only quantifier of entanglement \cite%
{Horodecki}. Whereas, for mixed states, entanglement can be discriminated
from separability by means of miscellaneous measures, which can be
distinguished to each other due to their operational meaning and their
mathematical properties \cite{adesso}. Using $\alpha =2$ in Eq. (\ref{E15}), we simply obtain $\mathcal{S}_{2}%
\mathcal{(\rho )=}-\ln \left[ \mathrm{Tr}\left( \rho ^{2}\right) \right] $,
which corresponds to the R\'{e}nyi-2 entropy \cite{ASG}. For an arbitrary
tripartite state $\rho _{ABC}$, it has been proven that the entropy $%
\mathcal{S}_{2}\mathcal{(\rho )}$ satisfies the so called strong
subadditivity inequality, i.e., $\mathcal{S}_{2}\left( \rho _{AB}\right) +%
\mathcal{S}_{2}\left( \rho _{BC}\right) \geqslant \mathcal{S}_{2}\left( \rho
_{ABC}\right) +\mathcal{S}_{2}\left( \rho _{B}\right) $ \cite{ASG}, which
allows to develop various measures of quantum correlations including
entanglement and quantum discord.

\subsection{ Gaussian R\'{e}nyi-2 entanglement}

For a bipartite Gaussian state with covariance matrix $\mathcal{V}_{AB}$ (%
\ref{E14}$)$, the Gaussian R\'{e}nyi-$2$ entanglement is defined as \cite%
{ASG}
\begin{equation}
\mathcal{E}_{\mathcal{R}}(\mathcal{V}_{AB}):=\underset{\{\sigma
_{_{xy}}|0<\sigma _{_{xy}}\leq \mathcal{V}_{AB},\text{ }\det \sigma
_{_{xy}}=1\}}{\inf }\mathcal{S}_{2}(\sigma _{x}),\text{ }  \label{E16}
\end{equation}%
where $\mathcal{S}_{2}(\varrho )=\frac{1}{2}\ln [\det \Theta ]$ is the
Gaussian R\'{e}nyi-2 entropy of the state $\varrho $ having the covariance
matrix $\Theta $ \cite{ASG}. In Eq. (\ref{E16}), the optimization is taken
over a pure two-mode Gaussian state with covariance matrix $\sigma _{_{xy}}$
smaller than $\mathcal{V}_{AB}$, with $\sigma _{x}$ being the marginal
covariance matrix of party $x$ obtained from $\sigma _{_{xy}}$ by partial
tracing over party $y$.

For generally mixed two-mode Gaussian states $\rho _{AB}$, Eq. (\ref{E16})
admits a complicated expression \cite{ASG}. In particular, for the two-mode
Gaussian squeezed thermal state $\rho _{AB}$ with the covariance matrix (\ref%
{E14}), Eq. (\ref{E16}) reads \cite{adesso}
\begin{equation}
\mathcal{E}_{\mathcal{R}}=\left\{
\begin{array}{c}
\frac{1}{2}\ln \left[ \frac{(g+1)s-\sqrt{\left[ (g-1)^{2}-4d^{2}\right] %
\left[ s^{2}-d^{2}-g\right] }}{4(d^{2}+g)^{2}}\right] ^{2}\text{if \ \ \ \ }%
2|d|+1\leq g<2s-1, \\
0\text{ \ \ \ \ \ \ \ \ \ \ \ \ \ \ \ \ \ \ \ \ \ \ \ \ \ \ \ \ \ \ \ \ \ \
\ \ \ \ \ \ \ \ \ \ \ \ \ \ \ \ \ \ \ \ \ \ \ \ \ \ \ \ \ if \ \ \ \ \ \ }%
g\geqslant 2s-1,%
\end{array}%
\right.  \label{E17}
\end{equation}%
where $s=\left( a+b\right) /2$, $d=\left( a-b\right) /2$ and $g=\sqrt{\det
\mathcal{V}_{AB}}$.

It is interesting to notice here that a comparative analysis performed in
\cite{JCTP}, between the Gaussian R\'{e}nyi-1 entanglement (that is the
entanglement of formation defined via the von Neumann entropy) and the
Gaussian R\'{e}nyi-2 entanglement given by Eq. (\ref{E17}), showed that the
former may not be the best choice for characterizing entanglement even when
restricted to the simple case of two-mode Gaussian states.

\subsection{Gaussian R\'{e}nyi-2 discord}

In the spirit of the measures proposed in \cite{Adesso}, where Gaussian
quantum discord has been defined via the conventional von Neumann entropy,
Adesso and the co-workers \cite{ASG} have obtained closed formula of the
Gaussian R\'{e}nyi-2 discord for generally mixed two-mode Gaussian states.

In a two-mode Gaussian state $\rho _{AB}$, the Gaussian R\'{e}nyi-2 discord $%
\mathcal{D}_{\mathcal{R}}$ is defined as the difference between the total
correlations, quantified by the Gaussian R\'{e}nyi-2 mutual information $%
\mathcal{I}_{\mathcal{R}}$, and the Gaussian R\'{e}nyi-2 classical
correlations $\mathcal{C}_{\mathcal{R}}$ \cite{Schumacher,Groisman}, i.e.,
\begin{equation}
\mathcal{D}_{\mathcal{R}}(\rho _{AB})\doteq \mathcal{I}_{\mathcal{R}}(\rho
_{AB})-\mathcal{C}_{\mathcal{R}}(\rho _{AB}),  \label{E18}
\end{equation}%
where $\mathcal{I}_{\mathcal{R}}(\rho _{AB})$ is defined by
\begin{eqnarray}
\mathcal{I}_{\mathcal{R}}(\rho _{AB}) &=&\mathcal{S}_{2}(\rho _{A})\mathcal{%
+S}_{2}(\rho _{B})\mathcal{-S}_{2}(\rho _{AB}),  \label{E19} \\
&=&\frac{1}{2}\ln \frac{\det \mathcal{V}_{A}\det \mathcal{V}_{B}}{\det
\mathcal{V}_{AB}}.  \label{E20}
\end{eqnarray}

From an operational point of view, the Gaussian R\'{e}nyi-2 mutual
information (\ref{E20}) quantifies the phase space distinguishability
between the Wigner function of the two-mode Gaussian state $\rho _{AB}$ and
the Wigner function of the product of the marginal $\rho _{A}\otimes \rho
_{B}$.

\begin{figure}[t]
\centerline{\includegraphics[width=0.45%
\columnwidth,height=5.3cm]{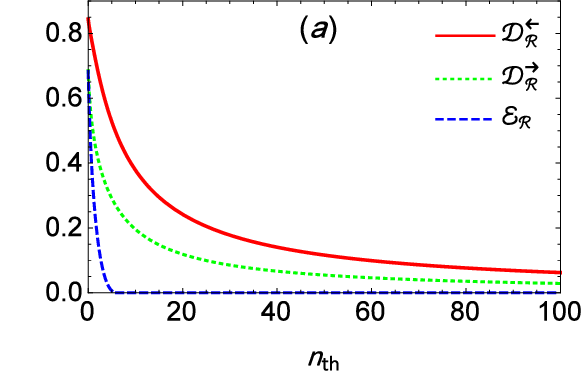}\includegraphics[width=0.45%
\columnwidth,height=5.5cm]{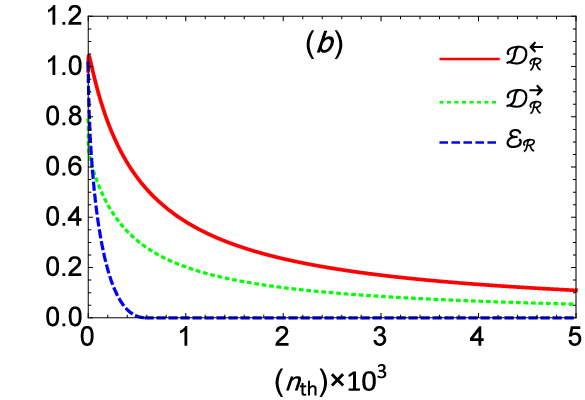}}
\caption{Plot of the Gaussian R\'{e}nyi-2 entanglement $\mathcal{E}_{%
\mathcal{R}}$ and the Gaussian R\'{e}nyi-2 discords $\mathcal{D}_{\mathcal{R}%
}^{\leftarrow}$ and $\mathcal{D}_{\mathcal{R}}^{\rightarrow}$ versus the
mean thermal phonon number $n_{\text{th}}$ using $\protect\eta=0.35$ and $%
\protect\kappa =3.85~\text{kHz}$. For the linear gain coefficient, we used $%
\mathcal{A}=200~\text{kHz}$ in panel (a) and $\mathcal{A}=20~\text{MHz}$ in
panel (b). As shown, $\mathcal{E}_{\mathcal{R}}$ is more degradable under
thermal noise, having a tendency to vanish rapidly. In contrast, the
discords $\mathcal{D}_{\mathcal{R}}^{\leftarrow}$ and $\mathcal{D}_{\mathcal{%
R}}^{\rightarrow}$ are more robust against thermal noise, where they are
significantly nonzero for $n_{\text{th}}>100$ in panel (a) and for $n_{\text{%
th}}>5\times10^{3}$ in panel (b). These observations entail that the
decoherence effect induced by high values of $n_{\text{th}}$ could be
overcome via preparing the atoms with a higher degree of quantum coherence,
i.e., $\mathcal{A}\gg1$ and $\protect\eta\rightarrow 0$.}
\label{Fig3}
\end{figure}

Following \cite{ASG,Schumacher,Groisman}, the Gaussian R\'{e}nyi-2 mutual
information $\mathcal{I}_{\mathcal{R}}(\rho _{AB})$ can be interpreted as
the degree of extra-discrete information that requires to be transmitted
among a continuous variable channel for reconstructing the complete joint
Wigner function of the two-mode Gaussian state $\rho _{AB}$ rather than just
the two marginal Wigner functions of the two modes $A$ and $B$. $\mathcal{I}%
_{\mathcal{R}}$ is always positive and vanishes only on product states,
i.e., $\rho _{AB}=\rho _{A}\otimes \rho _{B}$. Besides, the Gaussian R\'{e}nyi-2 classical correlations $\mathcal{C}_{%
\mathcal{R}}(\rho _{AB})$ are defined in terms of how much the ignorance
about the state of a party, saying $A$, is reduced when the most informative
local measurement is implemented on party $B$ \cite{Adesso}. Operationally,
we define $\mathcal{C}_{\mathcal{R}}^{\leftarrow }(\rho _{AB})(\mathcal{C}_{%
\mathcal{R}}^{\rightarrow }(\rho _{AB}))$ as the maximum decrease in the
Gaussian R\'{e}nyi-2 entropy of party $A(B)$, when a set of Gaussian local
measurements have been implemented on party $B(A)$, provided that the
optimisation is over all possible Gaussian measurements \cite{Adesso}
\begin{equation}
\mathcal{C}_{\mathcal{R}}^{\leftarrow }(\rho _{AB})=\underset{\Gamma _{B}^{M}%
}{\sup }\left[ \frac{1}{2}\ln \frac{\det \mathcal{V}_{A}}{\det \mathcal{%
\tilde{V}}_{A}^{M}}\right] ,  \label{E21}
\end{equation}%
where $\mathcal{\tilde{V}}_{A}^{M}=\mathcal{V}_{A}-\mathcal{V}_{A/B}(%
\mathcal{V}_{B}+\Gamma _{B}^{M})^{-1}\mathcal{V}_{A/B}^{\mathrm{T}}$ is the
covariance matrix of the mode $A$ after an optimised Gaussian measurement is
performed on mode $B$, with $\Gamma _{B}^{M}$ being a positive operator
valued measure \cite{Cirac}. $\mathcal{C}_{\mathcal{R}}^{\rightarrow }(\rho
_{AB})$ can be obtained from Eq. (\ref{E21}) by performing the double
exchange $\mathcal{V}_{A}\leftrightarrow \mathcal{V}_{B}$ and $\Gamma
_{B}^{M}\rightarrow \Gamma _{A}^{M}$. For the two-mode Gaussian states $\rho
_{AB}$ with covariance matrix (\ref{E14}), Eq. (\ref{E18}) reads \cite{ASG}
\begin{eqnarray}
\mathcal{D}_{\mathcal{R}}^{\leftarrow }(\rho _{AB}) &\doteq &\mathcal{I}_{%
\mathcal{R}}(\rho _{AB})-\mathcal{C}_{\mathcal{R}}^{\leftarrow }(\rho _{AB})%
\text{,}  \label{E22} \\
\mathcal{D}_{\mathcal{R}}^{\leftarrow }(\rho _{AB}) &=&\inf \left[ \frac{1}{2%
}\ln \frac{\det \mathcal{V}_{B}\det \mathcal{\tilde{V}}_{A}^{M}}{\det
\mathcal{V}_{AB}}\right] ,  \label{E23}
\end{eqnarray}%
with
\begin{equation}
\inf \left[ \ln \det \mathcal{\tilde{V}}_{A}^{M}\right] =\left\{
\begin{array}{c}
a\left( a-\frac{c^{2}}{b}\right) \text{\ if\ }\left[ ab^{2}c^{\prime
2}-c^{2}\left( a+bc^{\prime 2}\right) \right] \left[ ab^{2}c^{2}-c^{\prime
2}\left( a+bc^{2}\right) \right] <0, \\
\left[ \frac{\left\vert cc^{\prime }\right\vert +\sqrt{\left[ a\left(
b^{2}-1\right) -bc^{\prime }{}^{2}\right] \left[ a\left( b^{2}-1\right)
-bc^{2}\right] }}{b^{2}-1}\right] ^{2}\text{ otherwise}.%
\end{array}%
\right.  \label{E24}
\end{equation}

In general, the Gaussian R\'{e}nyi-2 discord is nonsymmetric by exchanging
the roles played by the modes $A$ and $B$, i.e., $\mathcal{D}_{\mathcal{R}%
}^{\leftarrow }\neq \mathcal{D}_{\mathcal{R}}^{\rightarrow }$, where the
discord $\mathcal{D}_{\mathcal{R}}^{\leftarrow }$($\mathcal{D}_{\mathcal{R}%
}^{\rightarrow }$) is obtained after Gaussian measurements have been
performed on mode $B$($A$). Moreover, form Eqs. (\ref{E23}) and (\ref{E24}),
we remark that if $c=0$ or $c^{\prime}=0$, the discords $\mathcal{D}_{%
\mathcal{R}}^{\leftarrow }$ and $\mathcal{D}_{\mathcal{R}}^{\rightarrow }$
vanish, which turns the state $\rho _{AB}$ into a product state with no
position or momentum correlations between the modes $A$ and $B$.

Notice that a comparative study \cite{Luo2} between the entanglement of
formation and quantum discord, in a two-qubit state, showed that these two
measures of quantum correlations are not only quantitatively but also
qualitatively different.

\begin{figure}[t]
\centerline{\includegraphics[width=0.45%
\columnwidth,height=5.5cm]{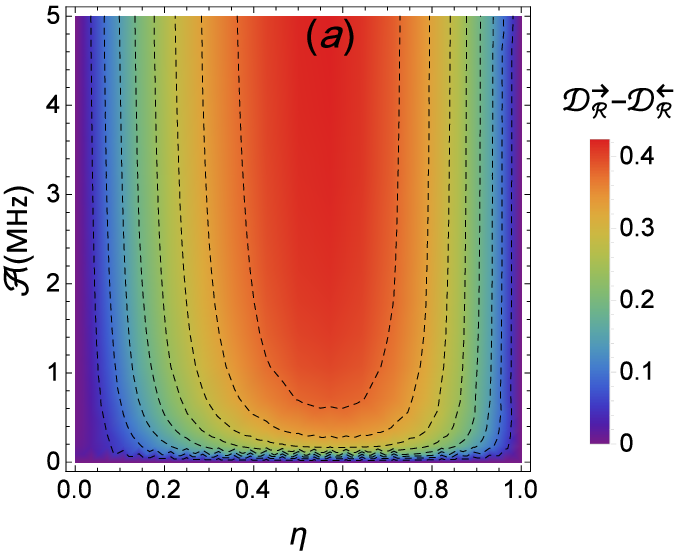}\includegraphics[width=0.45%
\columnwidth,height=5.5cm]{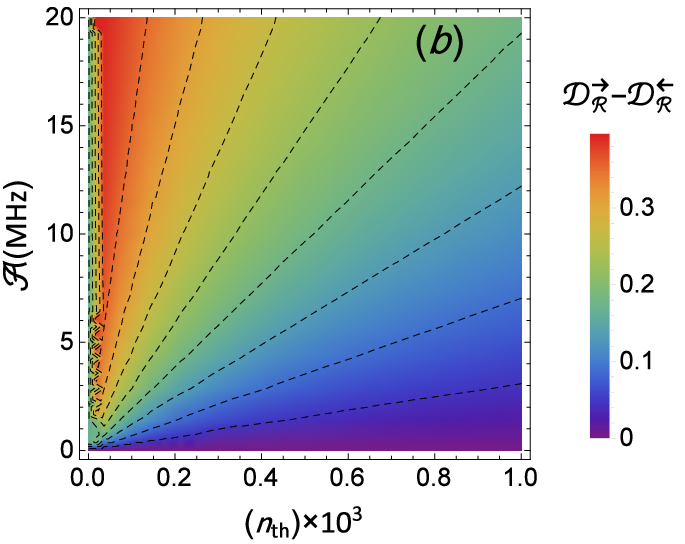}}
\caption{(a) Density plot of the difference $\mathcal{D}_{\mathcal{R}%
}^{\leftarrow}-\mathcal{D}_{\mathcal{R}}^{\rightarrow}$ against $\protect%
\eta $ and $\mathcal{A}$ using $\protect\kappa =3.85~\text{kHz}$ and $n_{%
\text{th}}=5$. (b) Density plot of the difference $\mathcal{D}_{\mathcal{R}%
}^{\leftarrow}-\mathcal{D}_{\mathcal{R}}^{\rightarrow}$ against $n_{\text{th}%
}$ and $\mathcal{A}$ using $\protect\kappa =3.85~\text{kHz}$ and $\protect%
\eta=0.35$. Under various conditions, we remark that $\mathcal{D}_{\mathcal{R%
}}^{\leftarrow}\geq\mathcal{D}_{\mathcal{R}}^{\rightarrow}$, meaning that
more quantumness of the studied three-level laser could be captured by
performing Gaussian measurements of mode $B$ that emitted during the second
transition $|l_{2}\rangle\rightarrow |l_{3}\rangle$.}
\label{Fig4}
\end{figure}

For realistic estimation of the Gaussian R\'{e}nyi-2 entanglement $\mathcal{E%
}_{\mathcal{R}}$ and the Gaussian R\'{e}nyi-2 discords $\mathcal{D}_{%
\mathcal{R}}^{\leftarrow}$ and $\mathcal{D}_{\mathcal{R}}^{\rightarrow}$ we
use parameters from \cite{xiong,Meschede}: the cavity decay rate $\kappa
=3.85~\text{kHz}$, the atomic decay rate $\gamma =20~\text{kHz}$, the
coupling constant $\upsilon =43~\text{kHz}$, the rate at which the atoms are
injected into the cavity $r=22~\text{kHz. Using these values, we get }%
\mathcal{A}=\frac{2r\upsilon ^{2}}{\gamma ^{2}}\approx 200~\text{kHz}$ for
the linear gain coefficient.

In Fig. \ref{Fig2}, we plot simultaneously $\mathcal{E}_{\mathcal{R}}$, $%
\mathcal{D}_{\mathcal{R}}^{\leftarrow }$ and $\mathcal{D}_{\mathcal{R}%
}^{\rightarrow }$ against the population inversion $\eta $ using $n_{\mathrm{%
th}}=5$ as value of the mean thermal phonon number. For the linear gain
coefficient, we used $\mathcal{A}=100~\text{kHz}$ in Fig. \ref{Fig2}a, $%
\mathcal{A}=1~\text{MHz}$ in Fig. \ref{Fig2}b, and $\mathcal{A}=50~\text{MHz}
$ in Fig. \ref{Fig2}c. Quite remarkably, for $\eta =1$, neither entanglement
nor discord in both directions could be detected ($\mathcal{E}_{\mathcal{R}}=%
\mathcal{D}_{\mathcal{R}}^{\leftarrow}=\mathcal{D}_{\mathcal{R}%
}^{\rightarrow}=0$) irrespective of the values of $\mathcal{A.}$ This can be
understood as follows: since $\eta =1$ corresponds to the situation in which
all the three-level atoms are populated in the ground level $|l_{3}\rangle $%
, then it will be no possibility for laser emission by the atoms via the
cascade transition $|l_{1}\rangle \rightarrow |l_{2}\rangle \rightarrow
|l_{3}\rangle $. Consequently, correlated laser modes in the cavity are not
expected. On the other hand, by preparing the atoms in a quantum coherent
superposition of the upper $|l_{1}\rangle $ and ground $|l_{3}\rangle $
levels, i.e., $\eta \neq 1$, quantum features of the cavity radiation of the
nondegenerate three-level laser could be well detected. In particular, when
the atoms are initially prepared in a maximum coherent superposition, i.e., $%
\eta =0$, we remark that, solely, the Gaussian R\'{e}nyi-2 discords $%
\mathcal{D}_{\mathcal{R}}^{\leftarrow}$ and $\mathcal{D}_{\mathcal{R}%
}^{\rightarrow}$ can capture the quantumness of correlations of the cavity
light.

Figure \ref{Fig2}a illustrates that, for $\mathcal{A}=100~\text{kHz}$, the
Gaussian R\'{e}nyi-2 entanglement $\mathcal{E}_{\mathcal{R}}$ is always zero
regardless of the values of $\eta $. In contrast, the Gaussian R\'{e}nyi-2
discords $\mathcal{D}_{\mathcal{R}}^{\leftarrow }$ and $\mathcal{D}_{%
\mathcal{R}}^{\rightarrow }$ are always nonzero for all values of the
population inversion $\eta $ except for $\eta \neq 1$. However, by
augmenting the values of the coefficient $\mathcal{A}$, we remark that, in
addition to quantum discord, the entanglement $\mathcal{E}_{\mathcal{R}}$
could be also detected. As clearly indicated in Figs. \ref{Fig2}b and c,
high values of $\mathcal{A}=2r\upsilon ^{2}/\gamma ^{2}$ allow to generate
maximum entanglement and discord provided that the three-level atoms are
initially prepared in a relatively stronger coherent superposition ($\eta
\rightarrow 0$). This entails that quantum correlations can be enhanced via
augmenting the rate $r$ at which the atoms are injected into the cavity or
augmenting the coupling strength $\upsilon $, or using the two mechanisms at
the same time. Manifestly, Fig. \ref{Fig2}c shows an interesting situation
in which the entanglement $\mathcal{E}_{\mathcal{R}}$ exceeds the quantum
discords $\mathcal{D}_{\mathcal{R}}^{\leftarrow }$ and $\mathcal{D}_{%
\mathcal{R}}^{\rightarrow }$ in contrary to the predictions based on the
assumption that entanglement is only a part of quantum discord \cite{Reid}.

The situation in which quantum discord is found less than entanglement
reminds us of the analogous feature observed within two-qubit states in \cite%
{Luo2}. This can be explained as follows: purely quantum correlations
captured by means of quantum discord often emerge as a consequence of
quantum coherence property \cite{Zurek}. Whereas, entangled states may
involve more than purely quantum correlations, that is, entangled states
usually carry classical ones \cite{Horodecki}. On the other hand, inspired
by the results \cite{Anderson,Groisman} postulated that classical
correlations should not be less than quantum ones. Therefore the case
corresponds to $\mathcal{E}_{\mathcal{R}}>\text{Max}[\mathcal{D}_{\mathcal{R}%
}^{\leftarrow },\mathcal{D}_{\mathcal{R}}^{\rightarrow }]$ can be viewed as
a situation in which entanglement is a certain mixture of purely quantum and
purely classical correlations. Overall, Fig. \ref{Fig2} shows that when the Gaussian R\'{e}nyi-2
entanglement $\mathcal{E}_{\mathcal{R}}$ failed to capture quantum
correlations in the state $\rho _{AB}$, the Gaussian R\'{e}nyi-2 discord can
capture them even when less coherence are injected in the cavity (small
values of the linear gain coefficient $\mathcal{A}$ with $\eta \rightarrow 1$%
).

Now, we focus our attention on the behavior of the entanglement $\mathcal{E}%
_{\mathcal{R}}$ and the discords $\mathcal{D}_{\mathcal{R}}^{\leftarrow }$
and $\mathcal{D}_{\mathcal{R}}^{\rightarrow }$ under influence of thermal
effect. In Fig. \ref{Fig3}, we plot simultaneously $\mathcal{E}_{\mathcal{R}%
} $, $\mathcal{D}_{\mathcal{R}}^{\leftarrow }$ and $\mathcal{D}_{\mathcal{R}%
}^{\rightarrow }$ against the mean thermal phonon number $n_{\mathrm{th}}$
using $\mathcal{A}=200~\text{kHz}$ in Fig. \ref{Fig3}a and $\mathcal{A}=20~%
\text{MHz}$ in Fig. \ref{Fig3}b. In both panels, we used $\eta =0.35$. As
shown, the three measures of nonclassicality $\mathcal{E}_{\mathcal{R}}$, $%
\mathcal{D}_{\mathcal{R}}^{\leftarrow }$ and $\mathcal{D}_{\mathcal{R}%
}^{\rightarrow }$ are maximum for $n_{\mathrm{th}}=0$ and decrease with
increasing $n_{\mathrm{th}}$. Obviously, the entanglement $\mathcal{E}_{\mathcal{R}}$ is found more
affected by thermal noise than quantum discord, having a tendency to vanish
quickly. However, by using high values of the linear gain coefficient $%
\mathcal{A}$, entangled states could be preserved over a wide range of $n_{%
\mathrm{th}}$. This means that by injecting more and more quantum coherence
into the cavity, it is possible to overcome the decoherence effect induced
by thermal noise. Besides, the discords $\mathcal{D}_{\mathcal{R}%
}^{\leftarrow}$ and $\mathcal{D}_{\mathcal{R}}^{\rightarrow}$ are found more
robust against thermal noise, where they decrease monotonically with
increasing $n_{\mathrm{th}}$, but never vanish.

Strikingly, the Gaussian R\'{e}nyi-2 discords $\mathcal{D}_{\mathcal{R}%
}^{\leftarrow }$ and $\mathcal{D}_{\mathcal{R}}^{\rightarrow }$ still
nonzero up to $n_{\mathrm{th}}=100$ using $\mathcal{A}=200~$kHz in Fig. \ref%
{Fig3}a and up to $n_{\mathrm{th}}=5\times 10^{3} $ using $\mathcal{A}=20~$%
MHz in Fig. \ref{Fig3}b. These results assert that the quantumness of
correlations in a nondegenerate three-level laser could be captured even in
high environmental temperature provided that the atoms are initially
prepared in the cavity with a high degree of quantum coherence, i.e., $%
\mathcal{A}\gg 1$ and $\eta \rightarrow 0$. Also, it is interestingly to observe from Fig. \ref{Fig3} that within
separable states ($\mathcal{E}_{\mathcal{R}}=0$), the discords $\mathcal{D}_{%
\mathcal{R}}^{\leftarrow }$ and $\mathcal{D}_{\mathcal{R}}^{\rightarrow }$
remain almost constant, reaching an asymptotic regime over a wide range of $%
n_{\mathrm{th}}$. Such behavior is commonly known as the freezing behavior
of quantum discord beyond entanglement, where an insightful physical
interpretation of this phenomenon is given in \cite{Modi,Bromly}.

It is not hard to remark from Figs. \ref{Fig2} and \ref{Fig3} that the two
discords $\mathcal{D}_{\mathcal{R}}^{\leftarrow }$ and $\mathcal{D}_{%
\mathcal{R}}^{\rightarrow }$ reveal rather different trends in similar
conditions. This means that inferring on the laser mode $A$ based on the
measurements implemented on the laser mode $B$ is completely different from
the reverse process, which is an important example of the role played by the
observer in quantum mechanics \cite{JDER}. Quite remarkably, in various
circumstances, the discord $\mathcal{D}_{\mathcal{R}}^{\rightarrow }$
obtained by performing Gaussian measurements on mode $A$, emitted during the
first transition $|l_{1}\rangle \rightarrow |l_{2}\rangle $, remains less or
equal to the discord $\mathcal{D}_{\mathcal{R}}^{\leftarrow }$ obtained by
performing Gaussian measurements on mode $B$, emitted during the second
transition $|l_{2}\rangle \rightarrow |l_{3}\rangle $. In what follows, we
show that such behavior is independent of the values of physical and
environmental parameters ($\eta $, $\kappa $, $\mathcal{A}$, $n_{\text{th}}$%
) of the state $\rho _{AB}$.

In Fig. \ref{Fig4}a, we plot $\mathcal{D}_{\mathcal{R}}^{\leftarrow }-%
\mathcal{D}_{\mathcal{R}}^{\rightarrow }$ using a density values of the
parameters $\eta $ and $\mathcal{A}$ for $n_{\text{th}}=5$ and $\kappa =3.85~%
\text{kHz}$. While, in Fig. \ref{Fig4}b we plot $\mathcal{D}_{\mathcal{R}%
}^{\leftarrow }-\mathcal{D}_{\mathcal{R}}^{\rightarrow }$ using a density
values of the parameters $n_{\text{th}}$ and $\mathcal{A}$ for $\eta =0.35$
and $\kappa =3.85~\text{kHz}$. As vividly illustrated the difference $%
\mathcal{D}_{\mathcal{R}}^{\leftarrow }-\mathcal{D}_{\mathcal{R}%
}^{\rightarrow }$ is always superior or equal to zero in various
circumstances, which implies that $\mathcal{D}_{\mathcal{R}}^{\leftarrow
}\geq \mathcal{D}_{\mathcal{R}}^{\rightarrow }$. Furthermore, employing Eqs.
(\ref{E14}) and (\ref{E23}), we obtain%
\begin{equation}
\mathcal{D}_{\mathcal{R}}^{\leftarrow }-\mathcal{D}_{\mathcal{R}%
}^{\rightarrow }=\ln \left[ \frac{ab+b}{ab+a}\frac{ab+a-c^{2}}{ab+b-c^{2}}%
\right] ,  \label{E25}
\end{equation}%
where the discord $\mathcal{D}_{\mathcal{R}}^{\rightarrow }$ can be obtained
from the expression of $\mathcal{D}_{\mathcal{R}}^{\leftarrow }$ by
performing the exchange $a\leftrightarrow b$. With some algebra, one can
show that the difference $\mathcal{D}_{\mathcal{R}}^{\leftarrow }-\mathcal{D}%
_{\mathcal{R}}^{\rightarrow }$ given by Eq. (\ref{E25}) and the difference $%
a-b$ have the same-sign. On the other hand, using the expressions of $a$ and
$b$ defined from the covariance matrix (\ref{E14}), we get
\begin{equation}
a-b=\langle \varsigma _{1}^{\dag }\varsigma _{1}\rangle -\langle \varsigma
_{2}^{\dag }\varsigma _{2}\rangle =\frac{\mathcal{A}\left( 1-\eta +2n_{%
\mathrm{th}}\right) }{2\left( \kappa +\mathcal{A}\eta \right) },  \label{e26}
\end{equation}
which is always positive or equal to zero since $\kappa $, $\mathcal{A}$ and
$n_{\text{th}}$ are positive, and $0\leqslant \eta \leqslant 1$. Therefore,
we conclude that the situation $\mathcal{D}_{\mathcal{R}}^{\leftarrow }\geq
\mathcal{D}_{\mathcal{R}}^{\rightarrow }$ is always fulfilled by the state $%
\rho _{AB}$, which asserts that more quantumness of correlations, in the
state $\rho _{AB}$, can be captured by performing Gaussian measurements on
the laser mode $B$ that emitted during the second transition $|l_{2}\rangle
\rightarrow |l_{3}\rangle $. Finally, with the cavity-quantum electrodynamics approaches \cite%
{Rempe1,QED1} and the strategy of homodyne measurement \cite{GS}, our
results may be verified experimentally.

\section{Conclusion}

\label{s4}

In a two-mode Gaussian state $\rho _{AB}$, coupled to a common two-mode
thermal bath, a comparative study between two indicators of nonclassicality
(entanglement and discord) is presented. The mode $A$($B$) is generated
within the first(second) transition of a nondegenerate three-level cascade
laser. The stationary covariance matrix of the state $\rho _{AB}$ is
evaluated within the good cavity limit and the linear-adiabatic
approximation. The Gaussian R\'{e}nyi-2 entanglement $\mathcal{E}_{\mathcal{R%
}}$ is used to quantify entanglement, while the Gaussian R\'{e}nyi-2
discords $\mathcal{D}_{\mathcal{R}}^{\leftarrow}$ and $\mathcal{D}_{\mathcal{%
R}}^{\rightarrow}$ are employed for capturing the quantumness of
correlations. It is found that both entanglement and discord could be generated and
enhanced via controlling the physical and environmental parameters of the
state $\rho _{AB}$. Optimal entanglement and discord can be achieved when
the atoms are initially prepared in a relatively strong coherent
superposition, i.e., $\mathcal{A}\gg1$ and $\eta\rightarrow 0$. Quite
remarkably, the entanglement $\mathcal{E}_{\mathcal{R}}$ is found more
fragile against thermal effect, suffering a sudden death-like behavior. In
contrast, quantum discord is found more robust, seeming to be captured in a
wide range of environmental temperatures. Numerical as well as analytical
analysis showed that the discord $\mathcal{D}_{\mathcal{R}}^{\leftarrow}$
remains always superior or equal to the discord $\mathcal{D}_{\mathcal{R}%
}^{\rightarrow}$, meaning that more quantumness of correlations can be
captured by performing Gaussian measurements on the mode $B$ that generated
during the second transition.

These results fairly indicate that, over lossy-noisy channels, nondegenerate
three-level lasers can be useful in implementing some quantum information
tasks, especially for that do need entanglement. Finally, it would be
interesting to investigate the conditions under which the quantum
correlations of the two laser modes $A$ and $B$ can be transferred to two
mechanical modes for generating, e.g., quantum steering between them. We
hope to report on this in a forthcoming work.

\section*{Funding sources}

This research did not receive any specific grant from funding agencies in
the public, commercial, or not-for-profit sectors.

\section*{Conflict of Interest}

The author declares no conflict of interest.

\section*{Data Availability Statement}

This manuscript has no associated data.

\end{document}